\documentstyle[aps]{revtex}

\begin{document}
\title{Massive Gauge Field Theory Without Higgs Mechanism\\
III. Illustration of Unitarity}
\author{Jun-Chen Su}
\address{Center for Theorectical Physics, Department of Physics, Jilin University,\\
Changchun 130023,\\
People's Republic of China}
\maketitle

\begin{abstract}
To illustrate the unitarity of the massive gauge field theory described in
the foregoing papers, we calculate the scattering amplitudes up to the
fourth order of perturbation by the optical theorem and the Landau-Cutkosky
rule. In the calculations, it is shown that for a given process, if all the
diagrams are taken into account, the contributions arising from the
unphysical intermediate states included in the longitudinal part of the
gauge boson propagator and in the ghost particle propagator are completely
cancelled out with each other in the S-matrix elements. Therefore, the
unitarity of the S-matrix is perfectly ensured.

\end{abstract}

\section{Introduction}

In the preceding paper called paper II, it was proved that the S-matrix
given by the QCD with massive gluons is independent of the gauge parameter.
However, the gauge-independence of S-matrix previously was not considered by
some people to be a sufficient condition of the unitarity of a theory.
Therefore, to demonstrate the unitarity, it is necessary to check whether
the contributions arising from unphysical intermediate states to the
S-matrix element written for a given process are cancelled in a perturbative
calculation . Historically, as mentioned in paper I$^{[1]}$, several attempts%
$^{[2]-[8]}$ of establishing the massive gauge field theory without Higgs
bosons were eventually negated. The reason for this partly is due to that
the theories were criticized to suffer from the difficulty of unitarity$%
^{[9]-[16]}$. Whether our theory is unitary in perturbative calculations?
That just is the question we want to answer in this paper.

To exhibit the unitarity, we choose to calculate two-gauge boson and
fermion-antifermion scattering amplitudes given in the perturbative
approximation up to the order of $g^4$. According to the optical theorem$%
^{[17-19]}$, we only need to compute the imaginary parts of these
amplitudes. The imaginary part of an amplitude can be evaluated by the
following formula$^{[17-19]}$%
\begin{equation}
2ImT_{ab}={\sum_c}T_{ac}T_{bc}^{*}  \eqnum{1.1}
\end{equation}
which was derived from the unitarity condition of S-matrix: $S$ $%
S^{+}=S^{+}S=1$ and the definition: $S=1+iT$. We would like to emphasize
that the above formula holds provided that the intermediate states \{$c$\}
form a complete set. This means that when we use this formula to evaluate
the imaginary part of an amplitude, we have to work in Feynman gauge. In
this gauge, the gauge boson propagator and the ghost particle one are given
in the form $^{[1]}$ 
\begin{equation}
iD_{\mu \nu }^{ab}(k)=\frac{-i\delta ^{ab}g_{\mu \nu }}{k^2-m^2+i\varepsilon 
},  \eqnum{1.2}
\end{equation}
and 
\begin{equation}
i\Delta ^{ab}(k)=\frac{-i\delta ^{ab}}{k^2-m^2+i\varepsilon }  \eqnum{1.3}
\end{equation}
where $m$ denotes the gauge boson mass.

In Eq. (1.2), the unit tensor $g_{\mu \nu }$ can be represented by the
completeness of the gauge boson intermediate states of polarization. 
\begin{equation}
g_{\mu \nu }={\sum_{\lambda =0}^3}e_\mu ^\lambda (k)e_{\lambda \nu
}(k)=P_{\mu \nu }(k)+Q_{\mu \nu }(k)  \eqnum{1.4}
\end{equation}
where $P_{\mu \nu }(k)$ and $Q_{\mu \nu }(k)$ are the transverse and
longitudinal projectors, respectively. On the mass-shell, they are expressed
as 
\begin{equation}
P_{\mu \nu }(k)=g_{\mu \nu }-k_\mu k_\nu /m^2,Q_{\mu \nu }(k)=k_\mu k_\nu
/m^2  \eqnum{1.5}
\end{equation}

It is noted here that in some previous works,$^{[14][15]}$ the Landau gauge
propagators were chosen at beginning to examine the unitarity through
calculation of the imaginary part of transition amplitudes. This procedure,
we think, is not reasonable and can not give a correct result in any case .
This is because that in the Landau gauge, the gauge boson propagator only
includes the transverse projector $P_{\mu \nu }(k)$ which does not represent
a complete set of the intermediate polarized states as seen from Eq. (1.4).
Usually, the right hand side (RHS) of Eq. (1.1) is calculated by using the
Landau-Cutkosky (L-C) rule$^{[17-19]}$. By this rule, the intermediate
propagators should be replaced by their imaginary parts 
\begin{equation}
Im(k_i^2-m^2+i\varepsilon )^{-1}=-\pi \delta (k_i^2-m^2)\theta (k_0) 
\eqnum{1.6}
\end{equation}

Utilizing the L-C rule to calculate the imaginary parts of the two-boson and
fermion-antifermion scattering amplitudes, we find, the unitarity of our
theory is no problems. A key point to achieve this conclusion is how to deal
with the loop diagram given by the gauge boson four-line vertex which was
considered to give no contribution to the S-matrix element in the previous
investigations$^{[3][12-15]}$. This diagram can be viewed as a limit of the
loop diagram formed by the gauge boson three-line vertices when one internal
line in the latter loop is shrunk into a point. In this way, we are able to
isolate from the former loop a term contributed from the unphysical
intermediate states which just guarantees the cancellation of the unphysical
amplitudes.

The rest of this paper is arranged as follows. In section 2, we sketch the
unitarity of the S-matrix elements of order $g^2$. In section 3, we describe
the calculations of the imaginary part of the two-gauge boson scattering
amplitude in the perturbative approximation of the order $g^4$ and show how
the unitarity is ensured. In section 4, the same thing will be done for the
fermion-antifermion scattering. The last section serves to make comments and
discussions. In Appendix, we will discuss the sign of imaginary parts of the
loop diagrams by a rigorous calculation.

\section{ Unitarity of the tree diagrams of order $g^2$}

For tree diagrams of order $g^2$, the unitarity of their transition
amplitudes is directly ensured by the on-mass shell condition. To illustrate
this point, we discuss the fermion-antifermion (say, quark-antiquark) and
two-gauge boson (say, two gluon ) scattering taking place in the S-channel
as shown in Figs. (1) and (2).

For the fermion-antifermion scattering, the S-matrix element may be written
as 
\begin{equation}
T_{fi}=ig^2j^\mu (p_1,p_2)D_{\mu \nu }(k)j^\nu (p_1^{\prime },p_2^{\prime })
\eqnum{2.1}
\end{equation}
where 
\begin{equation}
j^\mu (p_1,p_2)=\overline{v}(p_2)\frac{\lambda ^a}2\gamma ^\mu u(p_1) 
\eqnum{2.2}
\end{equation}
and 
\begin{equation}
D_{\mu \nu }(k)=\frac{g_{\mu \nu }-k_\mu k_\nu /k^2}{k^2-m^2+i\varepsilon }+%
\frac{\alpha k_\mu k_\nu /k^2}{k^2-\alpha m^2+i\varepsilon }.  \eqnum{2.3}
\end{equation}
Noticing $k=p_1+p_2=p_1^{\prime }+p_2^{\prime }$ and employing Dirac
equation, it is easy to see 
\begin{equation}
k_\mu j^\mu (p_1,p_2)=0  \eqnum{2.4}
\end{equation}
which holds due to that the theory only concerns the vector current and at
each vertex the fermion and antifermion have the same mass as in the QCD.
Therefore, the longitudinal term $k_\mu k_\nu /k^2$ in the propagator does
not contribute to the S-matrix in the approximation of order $g^2$. In other
words, the unphysical poles $k^2=0$ and $k^2=\alpha m^2$ do not appear in
the scattering amplitude.

For the process depicted in Fig. (2), the transition amplitude is 
\begin{equation}
\begin{array}{c}
T_{fi}=ig^2f^{abc}f^{a^{\prime }b^{\prime }c}e^\mu (k_1)e^\nu (k_2)e^{\mu
^{\prime }}(k_1^{\prime })^{*}e^{\nu ^{\prime }}(k_2^{\prime })^{*} \\ 
\times \Gamma _{\mu \nu \lambda }(k_1,k_2,q)D^{\lambda \lambda ^{\prime
}}(q)\Gamma _{\mu ^{\prime }\nu ^{\prime }\lambda ^{\prime }}(k_1^{\prime
},k_2^{\prime },q)
\end{array}
\eqnum{2.5}
\end{equation}
where 
\begin{equation}
\Gamma _{\mu \nu \lambda }(k_1,k_2,q)=g_{\mu \nu }(k_1-k_2)_\lambda +g_{\nu
\lambda }(k_2+q)_\mu -g_{\lambda \mu }(k_1+q)_\nu  \eqnum{2.6}
\end{equation}
and $e^\mu (k)$ stands for the gauge boson wave function satisfying the
transversity condition 
\begin{equation}
k_\mu e^\mu (k)=0.  \eqnum{2.7}
\end{equation}
The transversity of the polarized states and the relation $%
q=k_1+k_2=k_1^{\prime }+k_2^{\prime }$ directly lead to 
\begin{equation}
e^\mu (k_1)e^\nu (k_2)\Gamma _{\mu \nu \lambda }(k_1,k_2,q)q^\lambda =0. 
\eqnum{2.8}
\end{equation}
This equality, analogous to Eq. (2.4), guarantees the removal of the
unphysical poles from the S-matrix element written in Eq. (2.5).

Similarly, for the t-channel and u-channel diagrams, it is easy to verify
that the equalities in Eqs. (2.4) and (2.8) hold as well. These equalities
ensure the S-matrix elements for these diagrams and other processes such as
that a fermion and an antifermion annihilate into two bosons to be also
unitary.

The fact that the term $k_\mu k_\nu /k^2$ in the propagator gives no
contribution to the S-matrix elements means that the S-matrix is
gauge-independent at tree level. Therefore, in the gluon propagator, only
the physical pole at $k^2=m^2$ contributes to the S-matrix element and the
gluon propagator given in the Feynman gauge can reasonably be considered in
calculation of the S-matrix elements. Certainly, the fact mentioned above
allows us to write the intermediate states as transverse ones. When we
evaluate the imaginary part of the transition amplitudes by the L-C rule,
these intermediate states will be put on the mass shell.

\section{Unitarity of two-gluon scattering amplitude of order $g^4$}

In the preceding section. it was shown that in the lowest order
approximation of perturbation, the unitarity is no problem. How is it for
higher order perturbative approximations? To answer this question, in this
section, we investigate the unitarity of the two-gluon scattering amplitude
given in the order of $g^4$. For this purpose, we only need to consider the
diagrams shown in Figs. (3) and (4) and evaluate imaginary parts of the
amplitudes of these diagrams. The diagrams involving fermion intermediate
states are not necessarily taken into account because the fermion
intermediate state is already physical.

Fig. (3) contains eleven diagrams which have gauge boson intermediate
states. Except for the last diagram shown in Fig. (3k), the other diagrams
all have two-gauge boson intermediate states. If the unitarity condition is
satisfied, in the amplitudes given by these diagrams, the unphysical parts
arising from the longitudinally polarized intermediate states should be
cancelled by the amplitudes of the five diagrams depicted in Fig. (4) which
are of ghost intermediate states. To demonstrate this point, in the
following, we separately calculate the imaginary parts of the amplitudes of
all the diagrams in Figs. (3) and (4).

\subsection{The imaginary part of the diagrams in Figs.(3a)-(3j)}

By the L-C rule, the diagrams in Figs. (3a-3j) can be given by folding the
tree diagrams shown in Fig. (5) with their conjugates. Through the folding,
we obtain twice Figs. (3a-3f) and once Figs. (3g)-(3j). Noticing that each
of the diagrams in Figs. (3g)-(3j) has a symmetry factor $\frac 12$, we can
write the imaginary part of Figs. (3a)-(3j) as follows 
\begin{equation}
\begin{array}{c}
2ImT_1^{aba^{\prime }b^{\prime }}(p_1p_2;p_1^{\prime }p_2^{\prime })=\frac 12%
\int d\tau T_{\mu \nu }^{abcd}(p_1,p_2,k_1,k_2) \\ 
\times T_{\mu ^{\prime }\nu ^{\prime }}^{a^{\prime }b^{\prime
}cd*}(p_1^{\prime },p_2^{\prime };k_1,k_2)g^{\mu \mu ^{\prime }}g^{\nu \nu
^{\prime }}
\end{array}
\eqnum{3.1}
\end{equation}
where

\begin{equation}
\begin{array}{c}
d\tau =\frac{d^4k_1}{(2\pi )^4}\frac{d^4k_2}{(2\pi )^4}(2\pi )^4\delta
^4(p_1+p_2-k_1-k_2)\pi \delta (k_1^2-m^2)\theta (k_1^0) \\ 
\times \pi \delta (k_2^2-m^2)\theta (k_2^0)
\end{array}
\eqnum{3.2}
\end{equation}
and 
\begin{equation}
\begin{array}{c}
T_{\mu \nu }^{abcd}(p_1,p_2;k_1,k_2)=\sum\limits_{i=1}^4T_{\mu \nu
}^{(i)abcd}(p_1,p_2;k_1,k_2) \\ 
=-ig^2e^\rho (p_1)e^\sigma (p_{2)}\sum\limits_{i=1}^4T_{\rho \sigma \mu \nu
}^{(i)abcd}(p_1,p_2;k_1,k_2)
\end{array}
\eqnum{3.3}
\end{equation}
here $T_{\mu \nu }^{(i)abcd}(p_1,p_2;k_1,k_2)(i=1,2,3,4)$ stand for the
matrix elements of Figs. (5a)-(5d) respectively.

In light of Feynman rules, when we set 
\begin{equation}
C_1=f^{ace}f^{bde},C_2=f^{ade}f^{bce},C_3=f^{abe}f^{cde}  \eqnum{3.4}
\end{equation}
where the superscripts $abcd$ of $C_i$ have been suppressed for simplicity
and notice the relations given by the energy-momentum conservation 
\begin{equation}
q_1=p_1-k_1=k_2-p_2,\text{ }q_2=p_1-k_2=k_1-p_2,\text{ }q_3=p_1+p_2=k_1+k_2,
\eqnum{3.5}
\end{equation}
the functions $T_{\rho \sigma \mu \nu }^{(i)abcd}(p_1,p_2;k_1,k_2)$ may be
separately expressed as follows.

For Fig. (5a), 
\begin{equation}
\begin{array}{c}
T_{\rho \sigma \mu \nu }^{(1)abcd}(p_1,p_2;k_1,k_2)=\frac{C_1}{%
q_1^2-m^2+i\varepsilon }\Gamma _{\rho \mu \lambda }^{(1)}(p_1,k_1,q_1) \\ 
\times \Gamma _{\sigma \nu }^{(1)\lambda }(p_2,k_2,q_1)
\end{array}
\eqnum{3.6}
\end{equation}
where

\begin{equation}
\Gamma _{\rho \mu \lambda }^{(1)}(p_1,k_1,q_1)=g_{\rho \mu
}(k_1+p_1)_\lambda +g_{\mu \lambda }(q_1-k_1)_\rho -g_{\lambda \rho
}(q_1+p_1)_\mu ,  \eqnum{3.7}
\end{equation}
and 
\begin{equation}
\Gamma _{\sigma \nu \lambda }^{(1)}(p_2,k_2,q_1)=g_{\sigma \nu
}(p_2+k_2)_\lambda -g_{\nu \lambda }(q_1+k_2)_\sigma +g_{\lambda \sigma
}(q_1-p_2)_\nu .  \eqnum{3.8}
\end{equation}
For Fig. (5b), 
\begin{equation}
\begin{array}{c}
T_{\rho \sigma \mu \nu }^{(2)abcd}(p_1,p_2;k_1,k_2)=\frac{C_2}{%
q_2^2-m^2+i\varepsilon }\Gamma _{\rho \nu \lambda }^{(2)}(p_1,k_2,q_2)\times
\\ 
\Gamma _{\sigma \mu }^{(2)\lambda }(p_2,k_1,q_1)
\end{array}
\eqnum{3.9}
\end{equation}
where 
\begin{equation}
\Gamma _{\rho \nu \lambda }^{\left( 2\right) }(p_{1,}k_2,q_2)=g_{\rho \nu
}(p_1+k_2)_\lambda +g_{\nu \lambda }(q_2-k_2)_\rho -g_{\lambda \rho
}(q_2+p_1)_\nu  \eqnum{3.10}
\end{equation}
and 
\begin{equation}
\Gamma _{\sigma \mu \lambda }^{(2)}(p_2,k_1,q_2)=g_{\mu \sigma
}(p_2+k_1)_\lambda -g_{\mu \lambda }(k_1+q_2)_\sigma +g_{\lambda \sigma
}(q_2-p_2)_\mu .  \eqnum{3.11}
\end{equation}
For Fig. (5c), 
\begin{equation}
\begin{array}{c}
T_{\rho \sigma \mu \nu }^{(3)abcd}(p_1,p_2;k_1,k_2)=\frac{C_3}{%
q_3^2-m^2+i\varepsilon }\Gamma _{\rho \sigma \lambda }^{(3)}(p_1,p_2,q_3) \\ 
\times \Gamma _{\mu \nu }^{(3)\lambda }(k_1,k_2,q_3)
\end{array}
\eqnum{3.12}
\end{equation}
where 
\begin{equation}
\Gamma _{\rho \sigma \lambda }^{(3)}(p_1,p_2,q_3)=g_{\rho \sigma
}(p_1-p_2)_\lambda +g_{\sigma \lambda }(p_2+q_3)_\rho -g_{\lambda \rho
}(q_3+p_1)_\sigma  \eqnum{3.13}
\end{equation}
and 
\begin{equation}
\Gamma _{\mu \nu \lambda }^{(3)}(k_1,k_2,q_3)=g_{\mu \nu }(k_2-k_1)_\lambda
-g_{\nu \lambda }(k_2+q_3)_\mu +g_{\lambda \mu }(q_3+k_1)_\nu .  \eqnum{3.14}
\end{equation}
For Fig. (5d) 
\begin{equation}
T_{\rho \sigma \mu \nu }^{(4)abcd}(p_1,p_2;k_1,k_2)=C_1\gamma _{\rho \sigma
\mu \nu }^{(1)}+C_2\gamma _{\rho \sigma \mu \nu }^{(2)}+C_3\gamma _{\rho
\sigma \mu \nu }^{(3)}  \eqnum{3.15}
\end{equation}
where 
\begin{equation}
\gamma _{\rho \sigma \mu \nu }^{(1)}=g_{\rho \sigma }g_{\mu \nu }-g_{\sigma
\mu }g_{\rho \nu },  \eqnum{3.16}
\end{equation}
\begin{equation}
\gamma _{\rho \sigma \mu \nu }^{(2)}=g_{\rho \sigma }g_{\mu \nu }-g_{\rho
\mu }g_{\sigma \nu },  \eqnum{3.17}
\end{equation}
and 
\begin{equation}
\gamma _{\rho \sigma \mu \nu }^{(3)}=g_{\rho \mu }g_{\sigma \nu }-g_{\rho
\nu }g_{\sigma \mu }.  \eqnum{3.18}
\end{equation}
The expressions in Eqs. (3.1)-(3.3), (3.6), (3.9) and (3.12), as indicated
in Introduction, are all given in the Feynman gauge. When the intermediate
states $g^{\mu \mu ^{\prime }}$ and $g^{\nu \nu ^{\prime }}$ are decomposed
into the physical and unphysical parts in accordance with Eqs. (1.4) and
(1.5), Eq. (3.1) will be represented as 
\begin{equation}
\begin{array}{c}
2ImT_1=\frac 12\int d\tau T_{\mu \nu }^{abcd}T_{\mu ^{\prime }\nu ^{\prime
}}^{a^{\prime }b^{\prime }cd*}[P^{\mu \mu ^{\prime }}(k_1)P^{\nu \nu
^{\prime }}(k_2)+Q^{\mu \mu ^{\prime }}(k_1)g^{\nu \nu ^{\prime }} \\ 
+g^{\mu \mu ^{\prime }}Q^{\nu \nu ^{\prime }}(k_2)-Q^{\mu \mu ^{\prime
}}(k_1)Q^{\nu \nu ^{\prime }}(k_2)]
\end{array}
\eqnum{3.19}
\end{equation}
where $P^{\mu \mu ^{\prime }}(k_i)$ and $Q^{\mu \mu ^{\prime }}(k_i)$ are
defined in Eq. (1.5) and respectively represent the transverse (physical)
and longitudinal (unphysical) polarization intermediate states of gauge
bosons. We see, except for the first term, the other terms in Eq. (3,19) are
all related to the unphysical intermediate states. These terms should be
cancelled out in the total amplitude. In the following, we calculate these
terms separately. In the calculations, we note, the transversity of the
polarization vectors (see Eq. (2.7)), the relations written in Eq. (3.5) and
the on shell property of the momenta $p_1,p_2,k_1$ and $k_2$ will be often
used.

\subsubsection{Calculation of $T_{\mu \nu }^{abcd}T_{\mu ^{\prime }\nu
^{\prime }}^{a^{\prime }b^{\prime }cd*}Q^{\mu \mu ^{\prime }}(k_1)g^{\nu \nu
^{\prime }}$}

To calculate the second term in Eq. (3.19), according to the definition in
Eq. (3.3) for $T_{\mu \nu }^{abcd}$ and the expression in Eq. (1.5) for $%
Q^{\mu \mu ^{\prime }}(k_1)$, we need to calculate the contractions $k_1^\mu
T_{\rho \sigma \mu \nu }^{(i)abcd}(p_1,p_2;k_1,k_2)(i=1,2,3,4)$. From Eqs.
(3.7) and (3.8). we find 
\begin{equation}
k_1^\mu \Gamma _{\rho \mu \lambda }^{(1)}(p_1,k_1,q_1)=-q_{1\rho
}q_{1\lambda }+g_{\lambda \rho }(q_1^2-m^2)  \eqnum{3.20}
\end{equation}
and 
\begin{equation}
q_1^\lambda \Gamma _{\sigma \nu \lambda }^{(1)}(p_2,k_2,q_1)=-k_{2\nu
}q_{1\sigma }.  \eqnum{3.21}
\end{equation}
Using these equalities, from Eq. (3.6), we obtain 
\begin{equation}
k_1^\mu T_{\rho \sigma \mu \nu }^{(1)abcd}(p_1,p_2;k_1,k_2)=C_1k_{2\nu
}S_{\rho \sigma }^{(1)}(q_1)+C_1\Gamma _{\sigma \nu \rho }^{(1)}(p_2,k_2,q_1)
\eqnum{3.22}
\end{equation}
where 
\begin{equation}
S_{\rho \sigma }^{(1)}(q_1)=\frac{q_{1\rho }q_{1\sigma }}{%
q_1^2-m^2+i\varepsilon }.  \eqnum{3.23}
\end{equation}

Similarly, from Eqs. (3.10) and (3.11), one can get 
\begin{equation}
k_1^\mu \Gamma _{\sigma \mu \lambda }^{(2)}(p_2,k_1,q_2)=-q_{2\sigma
}q_{2\lambda }+g_{\lambda \sigma }(q_2^2-m^2)  \eqnum{3.24}
\end{equation}
and 
\begin{equation}
q_2^\lambda \Gamma _{\rho \nu \lambda }^{(2)}(p_1,k_2,q_2)=-q_{2\rho
}k_{2\nu }.  \eqnum{3.25}
\end{equation}
Based on these equalities, it is found form Eq. (3.9) 
\begin{equation}
k_1^\mu T_{\rho \sigma \mu \nu }^{(2)abcd}(p_1,p_2;k_1,k_2)=C_2k_{2\nu
}S_{\rho \sigma }^{(2)}(q_2)+C_2\Gamma _{\rho \nu \sigma }^{(2)}(p_1,k_2,q_2)
\eqnum{3.26}
\end{equation}
where 
\begin{equation}
S_{\rho \sigma }^{(2)}(q_2)=\frac{q_{2\rho }q_{2\sigma }}{%
q_2^2-m^2+i\varepsilon }  \eqnum{3.27}
\end{equation}

Along the same line, we can derive from Eqs. (3.13) and (3.14) that 
\begin{equation}
k_1^\mu \Gamma _{\mu \nu \lambda }^{(3)}(k_1,k_2,q_3)=k_{1\nu }q_{3\lambda
}+k_{1\lambda }k_{2\nu }-g_{\nu \lambda }(q_3^2-m^2)  \eqnum{3.28}
\end{equation}
and 
\begin{equation}
q_3^\lambda \Gamma _{\rho \sigma \lambda
}^{(3)}(p_1,p_2,q_3)=(k_1+k_2)^\lambda \Gamma _{\rho \sigma \lambda
}^{(3)}(p_1,p_2,q_3)=0,  \eqnum{3.29}
\end{equation}
thereby, we get from Eq. (3.12) 
\begin{equation}
k_1^\mu T_{\rho \sigma \mu \nu }^{(3)abcd}(p_1,p_2;k_1,k_2)=C_3k_{2\nu
}S_{\rho \sigma }^{(3)}(p_1,p_2,k_1)-C_3\Gamma _{\sigma \nu \rho
}^{(3)}(p_1,p_2,q_3)  \eqnum{3.30}
\end{equation}
where 
\begin{equation}
S_{\rho \sigma }^{(3)}(p_1,p_2,k_1)=\frac 1{q_3^2-m^2+i\varepsilon }%
k_1^\lambda \Gamma _{\rho \sigma \lambda }^{(3)}(p_1,p_2,q_3).  \eqnum{3.31}
\end{equation}

In addition, from Eq. (3.15), we may write 
\begin{equation}
\begin{array}{c}
k_1^\mu T_{\rho \sigma \mu \nu }^{(4)abcd}(p_1,p_2;k_1,k_2)=C_1\gamma _{\rho
\sigma \nu }^{(1)}(k_1)+C_2\gamma _{\rho \sigma \nu }^{(2)}(k_1) \\ 
+C_3\gamma _{\rho \sigma \nu }^{(3)}(k_1)
\end{array}
\eqnum{3.32}
\end{equation}
where 
\begin{equation}
\gamma _{\rho \sigma \nu }^{(1)}(k_1)=k_1^\mu \gamma _{\rho \sigma \mu \nu
}^{(1)}=g_{\rho \sigma }k_{1\nu }-g_{\rho \nu }k_{1\sigma },  \eqnum{3.33}
\end{equation}
\begin{equation}
\gamma _{\rho \sigma \nu }^{(2)}(k_1)=k_1^\mu \gamma _{\rho \sigma \mu \nu
}^{(2)}=g_{\rho \sigma }k_{1\nu }-g_{\sigma \nu }k_{1\rho }  \eqnum{3.34}
\end{equation}
and 
\begin{equation}
\gamma _{\rho \sigma \nu }^{(3)}(k_1)=k_1^\mu \gamma _{\rho \sigma \mu \nu
}^{(3)}=g_{\sigma \nu }k_{1\rho }-g_{\rho \nu }k_{1\sigma }.  \eqnum{3.35}
\end{equation}

Summing up the results denoted in Eqs. (3.22), (3.26), (3.30) and (3.32) and
noticing Eq. (3.3), we have 
\begin{equation}
k_1^\mu T_{\rho \sigma \mu \nu }^{abcd}(p_1,p_2;k_1,k_2)=-ig^2e^\rho
(p_1)e^\sigma (p_2)[k_{2\nu }S_{\rho \sigma }^{abcd}+G_{\rho \sigma \nu
}^{abcd}]  \eqnum{3.36}
\end{equation}
where 
\begin{equation}
S_{\rho \sigma }^{abcd}={\sum_{i=1}^3}C_iS_{\rho \sigma }^{(i)}  \eqnum{3.37}
\end{equation}
and 
\begin{equation}
G_{\rho \sigma \nu }^{abcd}={\sum_{i=1}^3}C_iG_{\rho \sigma \nu }^{(i)} 
\eqnum{3.38}
\end{equation}
in which 
\begin{equation}
G_{\rho \sigma \nu }^{(1)}=\Gamma _{\sigma \nu \rho
}^{(1)}(p_2,k_2,q_1)+\gamma _{\rho \sigma \nu }^{(1)}(k_1),  \eqnum{3.39}
\end{equation}
\begin{equation}
G_{\rho \sigma \nu }^{(2)}=\Gamma _{\rho \nu \sigma
}^{(2)}(p_1,k_2,q_2)+\gamma _{\rho \sigma \nu }^{(2)}(k_2)  \eqnum{3.40}
\end{equation}
and 
\begin{equation}
G_{\rho \sigma \nu }^{(3)}=-\Gamma _{\rho \sigma \nu
}^{(3)}(p_1,p_2,q_3)+\gamma _{\rho \sigma \nu }^{(3)}(p_1,p_2,k_1). 
\eqnum{3.41}
\end{equation}
Employing the expressions given in Eqs. (3.8), (3.10), (3.13), (3.29) and
(3.33)-(3.35) and considering the relations among the momenta as shown in
Eq. (3.5) and the on-shell condition of the external momenta, it is not
difficult to find the following relation 
\begin{equation}
G_{\rho \sigma \nu }^{(1)}=-G_{\rho \sigma \nu }^{(2)}=-G_{\rho \sigma \nu
}^{(3)}.  \eqnum{3.42}
\end{equation}

Now, let us look at the color factors. According to the expression 
\begin{equation}
f^{abe}f^{cde}=\frac 2N(\delta _{ac}\delta _{bd}-\delta _{ad}\delta
_{bc})+(d_{ace}d_{bde}-d_{bce}d_{ade})  \eqnum{3.43}
\end{equation}
and defining 
\begin{equation}
\begin{array}{c}
\beta _1=\frac 2N\delta _{ab}\delta _{cd}+d_{abe}d_{cde}, \\ 
\beta _2=\frac 2N\delta _{ad}\delta _{bc}+d_{ade}d_{bce}, \\ 
\beta _3=\frac 2N\delta _{ac}\delta _{bd}+d_{ace}d_{bde},
\end{array}
\eqnum{3.44}
\end{equation}
we may write 
\begin{equation}
C_1=\beta _1-\beta _2,C_2=\beta _1-\beta _3,C_3=\beta _3-\beta _2. 
\eqnum{3.45}
\end{equation}

Substitution of Eq. (3.45) into Eq. (3.38) and use of Eq. (3.42) directly
lead to 
\begin{equation}
\begin{array}{c}
G_{\rho \sigma \nu }^{abcd}=\beta _1(G_{\rho \sigma \nu }^{(1)}+G_{\rho
\sigma \nu }^{(2)})-\beta _2(G_{\rho \sigma \nu }^{(1)}+G_{\rho \sigma \nu
}^{(3)}) \\ 
+\beta _3(G_{\rho \sigma \nu }^{(3)}-G_{\rho \sigma \nu }^{(2)})=0.
\end{array}
\eqnum{3.46}
\end{equation}
This result makes Eq. (3.36) reduce to 
\begin{equation}
k_1^\mu T_{\mu \nu }^{abcd}=k_{2\nu }S^{abcd}  \eqnum{3.47}
\end{equation}
where

\begin{equation}
S^{abcd}=-ig^2e^\rho (p_1)e^\sigma (p_2)S_{\rho \sigma }^{abcd}. 
\eqnum{3.48}
\end{equation}
By using Eq. (3.47) and noticing $k_2^2=m^2,$ we finally obtain 
\begin{equation}
T_{\mu \nu }^{abcd}T_{\mu ^{\prime }\nu ^{\prime }}^{a^{\prime }b^{\prime
}cd^{*}}Q^{\mu \mu ^{\prime }}(k_1)g^{\nu \nu ^{\prime
}}=S^{abcd}S^{a^{\prime }b^{\prime }cd*}  \eqnum{3.49}
\end{equation}

It is emphasized that from the above derivation, we see, the four-line
vertex diagram in Fig. (5d) plays an essential role to give the relation in
Eq. (3.42) and hence to guarantee the cancellation of the second terms in
Eq. (3.22), (3.26) and (3.30), which are free from the pole at $q_i^2=m^2$,
as shown in Eq. (3.46).

\subsubsection{Calculation of $T_{\mu \nu }^{abcd}T_{\mu ^{\prime }\nu
^{\prime }}^{a^{\prime }b^{\prime }cd^{*}}g^{\mu \mu ^{\prime }}Q^{\nu \nu
^{\prime }}(k_2)$}

The procedure of calculating $T_{\mu \nu }^{abcd}T_{\mu ^{\prime }\nu
^{\prime }}^{a^{\prime }b^{\prime }cd^{*}}g^{\mu \mu ^{\prime }}Q^{\nu \nu
^{\prime }}(k_2)$ completely parallels to that described in the former
subsection. From Eqs. (3.7) and (3.8), it follows that 
\begin{equation}
k_2^\nu \Gamma _{\sigma \nu \lambda }^{(1)}(p_2,k_2,q_1)=-q_{1\sigma
}q_{1\lambda }+g_{\lambda \sigma }(q_1^2-m^2)  \eqnum{3.50}
\end{equation}
and 
\begin{equation}
q_1^\lambda \Gamma _{\rho \mu \lambda }^{(1)}(p_1,k_1,q_{1)}=-k_{1\mu
}q_{1\rho }.  \eqnum{3.51}
\end{equation}
These equalities allow us to get from Eq. (3.6) that 
\begin{equation}
k_2^\nu T_{\rho \sigma \mu \nu }^{(1)abcd}(p_1,p_2;k_1,k_2)=C_1k_{1\mu
}S_{\rho \sigma }^{(1)}+C_1\Gamma _{\rho \mu \sigma }^{(1)}(p_1,k_1,q_1). 
\eqnum{3.52}
\end{equation}
Based on the equalities 
\begin{equation}
k_2^\nu \Gamma _{\rho \nu \lambda }^{(2)}(p_1,k_2,q_2)=-q_{2\rho
}q_{2\lambda }+g_{\rho \lambda }(q_2^2-m^2)  \eqnum{3.53}
\end{equation}
and 
\begin{equation}
q_2^\lambda \Gamma _{\sigma \mu \lambda }^{(2)}(p_2,k_1,q_2)=-q_{2\sigma
}k_{1\mu }  \eqnum{3.54}
\end{equation}
which are derived from Eqs. (3.10) and (3.11), it is found 
\begin{equation}
k_2^\nu T_{\rho \sigma \mu \nu }^{(2)abcd}(p_1,p_2;k_1,k_2)=C_2k_{1\mu
}S_{\rho \sigma }^{(2)}+C_2\Gamma _{\sigma \mu \rho }^{(2)}(p_2,k_2,q_2). 
\eqnum{3.55}
\end{equation}
By making use of the equality 
\begin{equation}
k_2^\nu \Gamma _{\mu \nu \lambda }^{(3)}(k_1,k_2,q_3)=-k_{2\mu }q_{3\lambda
}-k_{1\mu }k_{2\lambda }+g_{\mu \lambda }(q_3^2-m^2)  \eqnum{3.56}
\end{equation}
which is derived from Eq. (3.14) and considering Eq. (3.29), we have 
\begin{equation}
k_2^\nu T_{\rho \sigma \mu \nu }^{(3)abcd}(p_1,p_2;k_1,k_2)=C_3k_{1\mu
}S_{\rho \sigma }^{(3)}+C_3\Gamma _{\rho \sigma \mu }^{(3)}(p_1,p_2,q_3). 
\eqnum{3.57}
\end{equation}
From Eq. (3.15), it is clear that 
\begin{equation}
\begin{array}{c}
k_2^\nu T_{\rho \sigma \mu \nu }^{(4)abcd}(p_1,p_2;k_1,k_2)=C_1\gamma _{\rho
\sigma \mu }^{(1)}(k_2) \\ 
+C_2\gamma _{\rho \sigma \mu }^{(2)}(k_2)+C_3\gamma _{\rho \sigma \mu
}^{(3)}(k_2)
\end{array}
\eqnum{3.58}
\end{equation}
where 
\begin{equation}
\gamma _{\rho \sigma \mu }^{(1)}(k_2)=k_2^\nu \gamma _{\rho \sigma \mu \nu
}^{(1)}=g_{\rho \sigma }k_{2\mu }-g_{\sigma \mu }k_{2\rho },  \eqnum{3.59}
\end{equation}
\begin{equation}
\gamma _{\rho \sigma \mu }^{(2)}{}(k_2)=k_2^\nu \gamma _{\rho \sigma \mu \nu
}^{(2)}=g_{\rho \sigma }k_{2\mu }-g_{\rho \mu }k_{2\sigma },  \eqnum{3.60}
\end{equation}
and 
\begin{equation}
\gamma _{\rho \sigma \mu }^{(3)}{}(k_2)=k_2^\nu \gamma _{\rho \sigma \mu \nu
}^{(3)}=g_{\rho \mu }k_{2\sigma }-g_{\sigma \mu }k_{2\rho }.  \eqnum{3.61}
\end{equation}
Combining Eqs. (3.52), (3.55), (3.57) and (3.58), we obtain 
\begin{equation}
k_2^\nu T_{\rho \sigma \mu \nu }^{abcd}(p_1,p_2;k_1,k_2)=-ig^2e^\rho
(p_1)e^\sigma (p_2)[k_{1\mu }S_{\rho \sigma }^{abcd}+\widetilde{G}_{\rho
\sigma \mu }^{abcd}]  \eqnum{3.62}
\end{equation}
where $S_{\rho \sigma }^{abcd}$ was defined in Eq. (3.37) and 
\begin{equation}
\widetilde{G}_{\rho \sigma \mu }^{abcd}(p_1,p_2;k_1,k_2)={\sum_{i=1}^3}C_i%
\widetilde{G}_{\rho \sigma \mu }^{(i)}  \eqnum{3.63}
\end{equation}
in which 
\begin{equation}
\widetilde{G}_{\rho \sigma \mu }^{(1)}=\Gamma _{\rho \mu \sigma
}^{(1)}(p_1,k_1,q_1)+\gamma _{\rho \sigma \mu }^{(1)}(k_2),  \eqnum{3.64}
\end{equation}
\begin{equation}
\widetilde{G}_{\rho \sigma \mu }^{(2)}=\Gamma _{\sigma \mu \rho
}^{(2)}(p_2,k_2,q_2)+\gamma _{\rho \sigma \mu }^{(3)}(k_2)  \eqnum{3.65}
\end{equation}
and 
\begin{equation}
\widetilde{G}_{\rho \sigma \mu }^{(3)}=\Gamma _{\rho \sigma \mu
}^{(3)}(p_1,p_2,q_3)+\gamma _{\rho \sigma \mu }^{(3)}(k_2).  \eqnum{3.66}
\end{equation}
Similar to Eq. (3.42), one may find 
\begin{equation}
\widetilde{G}_{\rho \sigma \mu }^{(1)}=-\widetilde{G}_{\rho \sigma \mu
}^{(2)}=-\widetilde{G}_{\rho \sigma \mu }^{(3)}.  \eqnum{3.67}
\end{equation}
These relations and those given in Eq. (3.45) also lead Eq. (3.63) to vanish 
\begin{equation}
\widetilde{G}_{\rho \sigma \mu }^{abcd}(p_1,p_2;k_1,k_2)=0.  \eqnum{3.68}
\end{equation}
Thus, Eq. (3.62) becomes 
\begin{equation}
k_2^\nu T_{\mu \nu }^{abcd}(p_1,p_2;k_1,k_2)=k_{1\mu }S^{abcd}  \eqnum{3.69}
\end{equation}
where $S^{abcd}$ was defined in Eq. (3.48) and thereby we have 
\begin{equation}
T_{\mu \nu }^{abcd}T_{\mu ^{\prime }\nu ^{\prime }}^{a^{\prime }b^{\prime
}cd*}g^{\mu \mu ^{\prime }}Q^{\nu \nu ^{\prime }}(k_2)=S^{abcd}S^{a^{\prime
}b^{\prime }cd*}.  \eqnum{3.70}
\end{equation}

\subsubsection{Calculation of $T_{\mu \nu }^{abcd}T_{\mu ^{\prime }\nu
^{\prime }}^{a^{\prime }b^{\prime }cd*}Q^{\mu \mu ^{\prime }}(k_1)Q^{\nu \nu
^{\prime }}(k_2)$}

To calculate $T_{\mu \nu }^{abcd}T_{\mu ^{\prime }\nu ^{\prime }}^{a^{\prime
}b^{\prime }cd*}Q^{\mu \mu ^{\prime }}(k_1)Q^{\nu \nu ^{\prime }}(k_2)$, it
is necessary to calculate $k_1^\mu k_2^\nu T_{\mu \nu }^{abcd}$. This may be
done in several ways. For example, we may simply contract Eq. (3.69) with
the vector $k_1^\mu $ to give 
\begin{equation}
k_1^\mu k_2^\nu T_{\mu \nu }^{abcd}(p_1,p_2;k_1,k_2)=m^2S^{abcd}. 
\eqnum{3.71}
\end{equation}
Certainly, paralleling to the procedure shown in the foregoing subsections,
we may first compute $k_1^\mu k_2^\nu T_{\rho \sigma \mu \nu }^{(i)abcd}$.
For instance, by contracting Eq. (3.56) with $k_1^\mu $, we derive 
\begin{equation}
k_1^\mu k_2^\nu \Gamma _{\mu \nu \lambda }^{(3)}(k_1,k_2,q_3)=-k_1\cdot
k_2q_{3\lambda }-m^2k_{2\lambda }+k_{1\lambda }(q_3^2-m^2).  \eqnum{3.72}
\end{equation}
From the above equality, noticing the identity in Eq. (3.29), it follows 
\begin{equation}
k_1^\mu k_2^\nu T_{\rho \sigma \mu \nu }^{(3)abcd}=C_3m^2S_{\rho \sigma
}^{(3)}+C_3k_1^\lambda \Gamma _{\rho \sigma \lambda }^{(3)}(p_1,p_2,q_3). 
\eqnum{3.73}
\end{equation}
The other terms can be given by contracting Eqs. (3.52), (3.55) and (3.58)
with $k_1^\mu $. Summing all these terms, one can exactly obtain the result
as written in Eq. (3.71). Employing Eq. (3.71), we get 
\begin{equation}
T_{\mu \nu }^{abcd}T_{\mu ^{\prime }\nu ^{\prime }}^{a^{\prime }b^{\prime
}cd*}Q^{\mu \mu ^{\prime }}(k_1)Q^{\nu \nu ^{\prime
}}(k_2)=S^{abcd}S^{a^{\prime }b^{\prime }cd*}.  \eqnum{3.74}
\end{equation}
Up to the present, the last three terms in Eq. (3.19) have been calculated.
Inserting Eqs. (3.49), (3.70) and (3.74) into Eq. (3.19), we arrive at 
\begin{equation}
\begin{array}{c}
2ImT_1^{aba^{\prime }b^{\prime }}=\frac 12\int d\tau T_{\mu \nu
}^{abcd}T_{\mu ^{\prime }\nu ^{\prime }}^{a^{\prime }b^{\prime }cd*}P^{\mu
\mu ^{\prime }}(k_1)P^{\nu \nu ^{\prime }}(k_2) \\ 
+\frac 12\int d\tau S^{abcd}S^{a^{\prime }b^{\prime }cd*}.
\end{array}
\eqnum{3.75}
\end{equation}
The second term in the above needs to be cancelled by the ghost diagrams.

\subsection{The imaginary part of the ghost diagrams in Fig. (4)}

The ghost diagrams in Fig. (4) can be given by folding the three tree
diagrams plotted in Fig. (6) with their conjugates. The folding gives double
Figs. (4a)-(4d) as well as Fig. (4e). Considering that the symmetry factor
of Fig. (4e) is $1$ other than $\frac 12$, the imaginary part of the
transition amplitude of Fig. (4) may be represented as 
\begin{equation}
\begin{array}{c}
2ImT_2^{aba^{\prime }b^{\prime }}=-\frac 12\int d\tau
T^{abcd}(p_1,p_2;k_1,k_2)T^{a^{\prime }b^{\prime }cd*}(p_1^{\prime
},p_2^{\prime };k_1,k_2) \\ 
-\frac 12\int d\tau T^{(3)abcd}(p_1,p_2;k_1,k_2)T^{(3)a^{\prime }b^{\prime
}cd*}(p_1^{\prime },p_2^{\prime };k_1,k_2)
\end{array}
\eqnum{3.76}
\end{equation}
where 
\begin{equation}
T^{abcd}(p_1,p_2;k_1,k_2)={\sum_{i=1}^3}T^{(i)abcd}(p_1,p_2;k_1,k_2) 
\eqnum{3.77}
\end{equation}
$T^{(i)abcd}(p_1,p_2;k_1,k_2)$ represent the matrix elements of Figs.
(6a)-(6c), and the minus sign is inherent for the ghost loops.

According to the Feynman rules and considering the transversity of the
polarization states, it is clear that 
\begin{equation}
T^{(i)abcd}(p_1,p_2;k_1,k_2)=S^{(i)abcd}  \eqnum{3.78}
\end{equation}
where 
\begin{equation}
S^{(i)abcd}=-ig^2e^\rho (p_1)e^\sigma (p_2)S_{\rho \sigma }^{(i)} 
\eqnum{3.79}
\end{equation}
here the $S_{\rho \sigma }^{(i)}(i=1,2,3)$ were defined in Eqs. (3.23),
(3.27) and (3.31) respectively. In accordance with Eq. (3.78), Eq. (3.76)
can be expressed as 
\begin{equation}
2ImT_2^{aba^{\prime }b^{\prime }}=-\frac 12\int d\tau S^{abcd}S^{a^{\prime
}b^{\prime }cd*}-\frac 12\int d\tau S^{(3)abcd}S^{(3)a^{\prime }b^{\prime
}cd*}  \eqnum{3.80}
\end{equation}
When adding Eq. (3.80) to Eq.(3.75), we see, the second term in Eq. (3.75)
is just cancelled by the first term in Eq. (3.80). However, still remains
the second term in Eq. (3.80) which represents half of the contribution of
the loop diagram in Fig. (4e) to the imaginary part of the amplitude, We are
particularly interested in the fact that the first term in Eq. (3.80)
contains the entire contributions from the ghost diagrams in Figs. (4a)-(4d)
and half of the contribution of the diagram in Fig. (4e). They completely
eliminate the unphysical part of the amplitudes given by Figs. (3a)-(3j),
needless to introduce any extra scalar particle for this elimination. How to
understand the remaining term in Eq. (3.80)? The occurrence of this term in
the sum of the amplitudes given in Eqs. (3.75) and (3.80) is due to that the
loop diagrams in Figs. (3g) and (4e) have different symmetry factors.
Therefore, only half of Fig. (4e) is needed to cancel the unphysical part of
Fig. (3g). It is reminded that until now, the loop diagram in Fig. (3k) has
not been considered. This diagram, as Fig. (3g), has also a symmetry factor $%
\frac 12$ and, as indicated soon later, gives a nonvanishing contribution to
the scattering amplitude and its imaginary part in the case of massive gauge
field theory. This contribution, of course, includes a part arising from the
unphysical intermediate states when we calculate the loop diagram in Fig.
(3k) in the Feynman gauge. Certainly, this unphysical part needs to be
cancelled by the corresponding ghost diagram as well. From the theoretical
logic, it is conceivable that the second term in Eq. (3.80) just serves such
a cancellation.

\subsection{The imaginary part of the diagram in Fig. (3k)}

How to evaluate the imaginary part of the amplitude of Fig. (3k) by the L-C
rule? This seems to be a difficult problem because we are not able to divide
the diagram into two parts by cutting the internal boson line of the closed
loop in Fig. (3k) without touching the vertex. However, we observe that when
letting one boson line of the closed loop in Fig. (3g) shrink into a point,
Fig. (3g) will be converted to Fig. (3k). This graphically intuitive
observation suggests that the amplitude given by Fig.(3k) can be treated as
a limit of the amplitude of Fig. (3g) when setting the momentum of one
propagator in the loop shown in Fig. (3g) tend to infinity. In this way, we
can isolate from Fig. (3k) the unphysical contribution which looks like to
be given by two-particle intermediate states and hence is able to compare
with the second term in Eq. (3.80). It is obvious that the difference
between the both diagrams in Figs. (3k) and (3g) only lies in their loops,
one of which is formed by the four-line vertex (See Fig. (7a)) and another
by the three-line vertex (see Fig. (7b)). Therefore, it is only necessary to
compare expressions of the two loops and establish a connection between them.

The expression of the loop in Fig. (7a) is 
\begin{equation}
\Pi _{\lambda \lambda ^{\prime }}^{(1)ab}(q)=-g^2f^{acd}f^{bcd}[g_{\lambda
\lambda ^{\prime }}g_{\mu \nu }-g_{\lambda \mu }g_{\lambda ^{\prime }\nu
}]\int \frac{d^4k}{(2\pi )^4}\frac{g^{\mu \nu }}{(k^2-m^2+i\varepsilon )}. 
\eqnum{3.81}
\end{equation}
The imaginary part of the above function have been exactly calculated in
Appendix. The result is 
\begin{equation}
Im\Pi _{\lambda \lambda ^{\prime }}^{(1)ab}(q)=-\frac{3g^2}{(4\pi )^2}%
f^{acd}f^{bcd}g_{\lambda \lambda ^{\prime }}\int_0^\infty \frac{dx}{x^2}%
sin(xm^2).  \eqnum{3.82}
\end{equation}
Clearly, it does not vanish when the mass $m$ is not equal to zero. The
above result is derived in the Feynman gauge, therefore, contains the
contribution arising from the unphysical intermediate states.

The expression of Fig. (7b) will be written in the form 
\begin{equation}
\Pi _{\lambda \lambda ^{\prime }}^{(2)ab}(q)=\int \frac{d^4k_1}{(2\pi )^4}%
\frac{d^4k_2}{(2\pi )^4}(2\pi )^4\delta ^4(k_1+k_2-q)\widetilde{\Pi }%
_{\lambda \lambda ^{\prime }}^{(2)ab}(k_1,k_2,q)  \eqnum{3.83}
\end{equation}
where 
\begin{equation}
\begin{array}{c}
\widetilde{\Pi }_{\lambda \lambda ^{\prime }}^{(2)ab}(k_1,k_2,q)=\frac 12%
g^2f^{acd}f^{bcd}\Gamma _{\mu \nu \lambda }(k_1,k_2,q) \\ 
\times \Gamma _{\mu ^{\prime }\nu ^{\prime }\lambda ^{\prime
}}(k_2,k_2,q)D^{\mu \mu ^{\prime }}(k_1)D^{\nu \nu ^{\prime }}(k_2)
\end{array}
\eqnum{3.84}
\end{equation}
in which the propagator $D^{\mu \nu }(k)$ was given in Eq. (2.3) with the
gauge parameter $\alpha =1$ and the vertex $\Gamma _{\mu \nu \lambda
}(k_1,k_2,q)$ was defined in Eq. (2.6). Let us take the limit :$\left|
k_{2\mu }\right| \rightarrow \infty $. In this limit, the product of the
propagator $D^{\nu \nu ^{\prime }}(k_2)$ and the vertices will approach to 
\begin{equation}
\begin{array}{c}
D^{\nu \nu ^{\prime }}(k_2)\Gamma _{\mu \nu \lambda }(k_1,k_2,q)\Gamma _{\mu
^{\prime }\nu ^{\prime }\lambda ^{\prime }}(k_1,k_2,q) \\ 
\rightarrow -\frac 1{k_2^2}g^{\nu \nu ^{\prime }}[g_{\mu \nu }k_{2\lambda
}-g_{\lambda \nu }k_{2\mu }][g_{\mu ^{\prime }\nu ^{\prime }}k_{2\lambda
^{\prime }}-g_{\lambda ^{\prime }\nu ^{\prime }}k_{2\mu ^{\prime }}] \\ 
=-\frac 1{k_2^2}[g_{\mu \mu ^{\prime }}k_{2\lambda }k_{2\lambda ^{\prime
}}+g_{\lambda \lambda ^{\prime }}k_{2\mu }k_{2\mu ^{\prime }}-g_{\mu \lambda
^{\prime }}k_{2\lambda }k_{2\mu ^{\prime }}-g_{\lambda \mu ^{\prime
}}k_{2\mu }k_{2\lambda ^{\prime }}].
\end{array}
\eqnum{3.85}
\end{equation}
If the tensor $k_{2\mu }k_{2\nu }/k_2^2$ behaves in such a way in the limit 
\begin{equation}
k_{2\mu }k_{2\nu }/k_2^2\rightarrow g_{\mu \nu }  \eqnum{3.86}
\end{equation}
(This limit will be justified in Appendix and a similar limitation for the
polarization vector was proposed in the proof of $\gamma _5$-anomaly (see
Ref. (20), Chapter 19)), then we find 
\begin{equation}
\widetilde{\Pi }_{\lambda \lambda ^{\prime }}^{(2)ab}(k_1,k_2,q)\text{ }_{%
\overrightarrow{\left| k_{2\mu }\right| \rightarrow \infty }\text{ }%
}g^2f^{acd}f^{bcd}(g_{\lambda \lambda ^{\prime }}g_{\mu \nu }-g_{\lambda \mu
}g_{\lambda ^{\prime }\nu })\frac{g^{\mu \nu }}{k_1^2-m^2+i\varepsilon } 
\eqnum{3.87}
\end{equation}
and hence 
\begin{equation}
\left| \Pi _{\lambda \lambda ^{\prime }}^{(2)ab}(q)\right| \text{ }_{%
\overrightarrow{\left| k_{2\mu }\right| \rightarrow \infty }\text{ }}\left|
\Pi _{\lambda \lambda ^{\prime }}^{(1)ab}(q)\right| .  \eqnum{3.88}
\end{equation}
Particularly, in the physical region, the sign of the imaginary part of the
amplitude $\Pi _{\lambda \lambda ^{\prime }}^{(2)ab}(q)$ is the same as the
corresponding part of the amplitude $\Pi _{\lambda \lambda ^{\prime
}}^{(1)ab}(q)$, as will be demonstrated in Appendix. In view of the argument
given above, the imaginary part of Fig. (3k) may equivalently be replaced by
the imaginary part of Fig. (3g) in the limit $\left| k_{2\mu }\right|
\rightarrow \infty $. Thus, we can write 
\begin{equation}
\begin{array}{c}
2ImT_3^{aba^{\prime }b^{\prime }}=\frac 12\int d\tau {\lim_{\left| k_{2\mu
}\right| \rightarrow \infty }}T_{\mu \nu }^{(3)abcd}T_{\mu ^{\prime }\nu
^{\prime }}^{(3)a^{\prime }b^{\prime }cd*}[P^{\mu \mu ^{\prime }}(k_1)P^{\nu
\nu ^{\prime }}(k_2) \\ 
+Q^{\mu \mu ^{\prime }}(k_1)g^{\nu \nu ^{\prime }}+g^{\mu \mu ^{\prime
}}Q^{\nu \nu ^{\prime }}(k_2)-Q^{\mu \mu ^{\prime }}(k_1)Q^{\nu \nu ^{\prime
}}(k_2)].
\end{array}
\eqnum{3.89}
\end{equation}
The first term in the above only concerns the physical intermediate states.
We do not pursue here what the limit for this term looks like because it is
of no importance at present. we are interested in examining the other three
terms. Look at the expression given in Eq. (3.28). The first term in it can
be ignored due to the equality in Eq. (3.29). The last term can also be
neglected comparing to the second term in the limit $\left| k_{2\mu }\right|
\rightarrow \infty $. Thus, Eq. (3.30) will be reduced to 
\begin{equation}
k_1^\mu T_{\rho \sigma \mu \nu }^{(3)abcd}\rightarrow C_3k_{2\nu }S_{\rho
\sigma }^{(3)}  \eqnum{3.90}
\end{equation}
where $S_{\rho \sigma }^{(3)}$ was defined in Eq. (3.31) and is irrelevant
to $k_2$. The result in Eq. (3.90) enables us to write the second term in
Eq. (3.89) as 
\begin{equation}
{\lim_{\left| k_{2\mu }\right| \rightarrow \infty }}T_{\mu \nu
}^{(3)abcd}T_{\mu ^{\prime }\nu ^{\prime }}^{(3)a^{\prime }b^{\prime
}cd*}Q^{\mu \mu ^{\prime }}(k_1)g^{\nu \nu ^{\prime
}}=S^{(3)abcd}S^{(3)a^{\prime }b^{\prime }cd*}  \eqnum{3.91}
\end{equation}
where $S^{(3)abcd}$ was defined in Eq. (3.79). In the above, the
compatibility of the on-shell condition $k_2^2=m^2$ with the limit $\left|
k_{2\mu }\right| \rightarrow \infty $ has been noticed .

By the same reason as stated above, only the second term in Eq. (3.56)
should be considered in the limit. Therefore, Eq. (3.57) is approximated to 
\begin{equation}
k_2^\nu T_{\rho \sigma \mu \nu }^{\left( 3\right) abcd}\rightarrow
-C_3k_{1\mu }\widetilde{S}_{\rho \sigma }^{(3)}  \eqnum{3.92}
\end{equation}
where 
\begin{equation}
\widetilde{S}_{\rho \sigma }^{(3)}={\lim_{\left| k_{2\mu }\right|
\rightarrow \infty }}\frac{k_2^\lambda \Gamma _{\rho \sigma \lambda
}^{(3)}(p_1,p_2;q_3)}{q_3^2-m^2+i\varepsilon }.  \eqnum{3.93}
\end{equation}
With the result in Eq. (3.92), the third term in Eq. (3.89) becomes 
\begin{equation}
{\lim_{\left| k_{2\mu }\right| \rightarrow \infty }}T_{\mu \nu
}^{(3)abcd}T_{\mu ^{\prime }\nu ^{\prime }}^{(3)a^{\prime }b^{\prime
}cd*}g^{\mu \mu ^{\prime }}Q^{\nu \nu ^{\prime }}(k_2)=\widetilde{S}%
^{(3)abcd}\widetilde{S}^{(3)a^{\prime }b^{\prime }cd^{*}}  \eqnum{3.94}
\end{equation}
where 
\begin{equation}
\widetilde{S}^{(3)abcd}=-ig^2e^\rho (p_1)e^\sigma (p_2)C_3\widetilde{S}%
_{\rho \sigma }^{(3)}.  \eqnum{3.95}
\end{equation}
Similarly, in the limit $\left| k_{2\mu }\right| \rightarrow \infty $, we
can neglect the first term (due to Eq. (3.29)) and the last term in Eq.
(3.72). The second term in Eq. (3.72) permits us to rewrite Eq. (3.73) in
the form 
\begin{equation}
k_1^\mu k_2^\nu T_{\rho \sigma \mu \nu }^{(3)abcd}\rightarrow -C_3m^2%
\widetilde{S}_{\rho \sigma }^{(3)}  \eqnum{3.96}
\end{equation}
which may more directly be given by contracting Eq. (3.92) with $k_1^\mu $.
From this result, it is clear to see 
\begin{equation}
{\lim_{\left| k_{2\mu }\right| \rightarrow \infty }}T_{\mu \nu
}^{(3)abcd}T_{\mu ^{\prime }\nu ^{\prime }}^{(3)a^{\prime }b^{\prime
}cd*}Q^{\mu \mu ^{\prime }}(k_1)Q^{\nu \nu ^{\prime }}(k_2)=\widetilde{S}%
^{(3)abcd}\widetilde{S}^{(3)a^{\prime }b^{\prime }cd*}.  \eqnum{3.97}
\end{equation}

On inserting Eqs. (3.91), (3.94) and (3.97) into Eq. (3.89), we see, the
last two terms in Eq. (3.89) are cancelled with each other. As a result, we
have 
\begin{equation}
\begin{array}{c}
2ImT_3^{aba^{\prime }b^{\prime }}=\frac 12\int d\tau {\lim_{\left| k_{2\mu
}\right| \rightarrow \infty }}T_{\mu \nu }^{(3)abcd}T_{\mu ^{\prime }\nu
^{\prime }}^{(3)a^{\prime }b^{\prime }cd*}P^{\mu \mu ^{\prime }}(k_1)P^{\nu
\nu ^{\prime }}(k_2) \\ 
+\frac 12\int d\tau S^{(3)abcd}S^{(3)a^{\prime }b^{\prime }cd*}.
\end{array}
\eqnum{3.98}
\end{equation}
The second term above is just cancelled by the second term in Eq. (3.80).
Combining the results given in Eqs. (3.75),(3.80) and (3.98), we obtain the
total amplitude as follows 
\begin{equation}
\begin{array}{c}
2ImT^{aba^{\prime }b^{\prime }}=\sum\limits_{i=1}^32ImT_i^{aba^{\prime
}b^{\prime }}=\frac 12\int d\tau T_{\mu \nu }^{abcd}T_{\mu ^{\prime }\nu
^{\prime }}^{a^{\prime }b^{\prime }cd*}P^{\mu \mu ^{\prime }}(k_1)P^{\nu \nu
^{\prime }}(k_2) \\ 
+\frac 12\int d\tau {\lim_{\left| k_{2\mu }\right| \rightarrow \infty }}%
T_{\mu \nu }^{(3)abcd}T_{\mu ^{\prime }\nu ^{\prime }}^{(3)a^{\prime
}b^{\prime }cd*}P^{\mu \mu ^{\prime }}(k_1)P^{\nu \nu ^{\prime }}(k_2)
\end{array}
\eqnum{3.99}
\end{equation}
which is only related to the physical intermediate states. Thus, the proof
of the unitarity is accomplished.

We note here that the results given in this subsection rely on how to
correctly treat the limit procedure. As will be shown in Appendix, the limit
given in Eq. (3.86) is the only choice of converting Fig. (3g) into Fig.
(3k) when the relation in Eq. (3.29) is considered. Similarly, to obtain the
desirable limiting results presented in Eqs. (3.91), (3.94) and (3.97) , the
reasonable expressions in Eqs. (3.28), (3.56) and (3.72) are necessary to be
used.

\section{Unitarity of fermion-antifermion scattering amplitude of order $g^4$%
}

In this section, to illustrate the unitarity of the theory further, we
evaluate the imaginary part of the fermion-antifermion scattering amplitude
in the perturbative approximation of order $g^4$. For this purpose, it is
only necessary to consider the diagrams shown in Fig. (8).

The diagrams in Figs. (8a)-(8e) can be reconstructed by folding the tree
diagrams in Figs. (9a)-(9c) with their conjugates. Since the folding gives
two times of Figs. (8a)-(8d) and one time of Fig. (8e) which possesses a
symmetry factor $\frac 12$, the imaginary parts of the amplitudes given by
Figs. (8a)-(8e) may be represented as 
\begin{equation}
\begin{array}{c}
2ImT_1=\frac 12\int d\tau T_{\mu \nu }^{ab}T_{\mu ^{\prime }\nu ^{\prime
}}^{ab*}g^{\mu \mu ^{\prime }}g^{\nu \nu ^{\prime }} \\ 
=\frac 12\int d\tau T_{\mu \nu }^{ab}T_{\mu ^{\prime }\nu ^{\prime
}}^{ab*}[P^{\mu \mu ^{\prime }}(k_1)P^{\nu \nu ^{\prime }}(k_2)+Q^{\mu \mu
^{\prime }}(k_1)g^{\nu \nu ^{\prime }} \\ 
+g^{\mu \mu ^{\prime }}Q^{\nu \nu ^{\prime }}(k_2)-Q^{\mu \mu ^{\prime
}}(k_1)Q^{\nu \nu ^{\prime }}(k_2)]
\end{array}
\eqnum{4.1}
\end{equation}
where 
\begin{equation}
T_{\mu \nu }^{ab}(p_1,p_2;k_1,k_2)={\sum_{i=1}^3}T_{\mu \nu
}^{(i)ab}(p_1,p_2;k_1,k_2)  \eqnum{4.2}
\end{equation}
$T_{\mu \nu }^{(i)ab}(p_1,p_2;k_1,k_2)$ ($i=1,2,3$) denote the matrix
elements of Figs. (9a)-(9c) respectively. According to the Feynman rules,
they can be written as 
\begin{equation}
T_{\mu \nu }^{(1)ab}(p_1,p_2;k_1,k_2)=-ig^2\overline{v}(p_2)\frac{\lambda ^b}%
2\gamma _\nu \frac{{\bf p}_1-{\bf k}_1+M}{(p_1-k_1)^2-M^2+i\varepsilon }%
\frac{\lambda ^a}2\gamma _\mu u(p_1),  \eqnum{4.3}
\end{equation}
\begin{equation}
T_{\mu \nu }^{(2)ab}(p_1,p_2;k_1,k_2)=-ig^2\overline{v}(p_2)\frac{\lambda ^a}%
2\gamma _\mu \frac{{\bf k}_1-{\bf p}_2+M}{(k_1-p_2)^2-M^2+i\varepsilon }%
\frac{\lambda ^b}2\gamma _\nu u(p_1)  \eqnum{4.4}
\end{equation}
and 
\begin{equation}
T_{\mu \nu }^{(3)ab}(p_1,p_2;k_1,k_2)=-\frac{g^2f^{abc}}{q^2-m^2+i%
\varepsilon }\Gamma _{\mu \nu \lambda }(k_1,k_2,q)\overline{v}(p_2)\frac{%
\lambda ^c}2\gamma ^\lambda u(p_1)  \eqnum{4.5}
\end{equation}
where $\Gamma _{\mu \nu \lambda }(k_1,k_2,q)$ was defined in Eq. (2.6), $M$
is the fermion mass and ${\bf p=}p^\mu \gamma _\mu $.

For evaluating the second term in Eq. (4.1), we need to compute the
contraction of $T_{\mu \nu }^{(i)ab}$ with $k_1^\mu $. By applying Dirac
equation, the on-mass shell condition and the relation $q=k_1+k_2=p_1+p_2$,
one may get 
\begin{equation}
\begin{array}{c}
k_1^\mu [T_{\mu \nu }^{(1)ab}+T_{\mu \nu }^{(2)ab}]=-ig^2\overline{v}(p_2)[%
\frac{\lambda ^a}2,\frac{\lambda ^b}2]\gamma _\nu u(p_1) \\ 
=g^2f^{abc}\overline{v}(p_2)\frac{\lambda ^c}2\gamma _\nu u(p_1)
\end{array}
\eqnum{4.6}
\end{equation}
and 
\begin{equation}
k_1^\mu T_{\mu \nu }^{(3)ab}=-g^2f^{abc}\overline{v}(p_2)\frac{\lambda ^c}2%
\gamma _\nu u(p_1)+k_{2\nu }S^{ab}.  \eqnum{4.7}
\end{equation}
where 
\begin{equation}
S^{ab}=\frac{g^2f^{abc}}{2p_1\cdot p_2+m^2}\overline{v}(p_2)\frac{\lambda ^c}%
2{\bf k}_1u(p_1)  \eqnum{4.8}
\end{equation}
Adding Eq. (4.7) to Eq. (4.6), we find 
\begin{equation}
k_1^\mu T_{\mu \nu }^{ab}=k_{2\nu }S^{ab}  \eqnum{4.9}
\end{equation}
As seen from the above, there is a cancellation among the diagrams in Figs.
(9a)-(9c). From Eq. (4.9), one may derive 
\begin{equation}
T_{\mu \nu }^{ab}T_{\mu ^{\prime }\nu ^{\prime }}^{ab*}Q^{\mu \mu ^{\prime
}}(k_1)g^{\nu \nu ^{\prime }}=S^{ab}S^{ab*}.  \eqnum{4.10}
\end{equation}

Let us turn to calculate the third term in Eq. (4.1). Along the same line
stated above, one may get 
\begin{equation}
k_2^\mu [T_{\mu \nu }^{(1)ab}+T_{\mu \nu }^{(2)ab}]=-g^2f^{abc}\overline{v}%
(p_2)\frac{\lambda ^c}2\gamma _\mu u(p_1)  \eqnum{4.11}
\end{equation}
and 
\begin{equation}
k_2^\nu T_{\mu \nu }^{(3)ab}=g^2f^{abc}\overline{v}(p_2)\frac{\lambda ^c}2%
\gamma _\mu u(p_1)+k_{1\mu }\widetilde{S}^{ab}  \eqnum{4.12}
\end{equation}
where 
\begin{equation}
\widetilde{S}^{ab}=\frac{-g^2f^{abc}}{q^2-m^2+i\varepsilon }\overline{v}(p_2)%
\frac{\lambda ^c}2{\bf k}_2u(p_1).  \eqnum{4.13}
\end{equation}
From the equality 
\begin{equation}
\overline{v}(p_2)\frac{\lambda ^c}2({\bf k}_1+{\bf k}_2)u(p_1)=\overline{v}%
(p_2)\frac{\lambda ^c}2({\bf p}_1+{\bf p}_2)u(p_1)=0,  \eqnum{4.14}
\end{equation}
it follows that 
\begin{equation}
\widetilde{S}^{ab}=S^{ab}.  \eqnum{4.15}
\end{equation}
Adding Eq. (4.12) to Eq. (4.11) and noticing Eq. (4.15), we have 
\begin{equation}
k_2^\nu T_{\mu \nu }^{ab}=k_{1\mu }S^{ab}.  \eqnum{4.16}
\end{equation}
This result gives rise to 
\begin{equation}
T_{\mu \nu }^{ab}T_{\mu ^{\prime }\nu ^{\prime }}^{ab*}g^{\mu \mu ^{\prime
}}Q^{\nu \nu ^{\prime }}(k_2)=S^{ab}S^{ab*}  \eqnum{4.17}
\end{equation}

For evaluating the last term in Eq. (4.1), we may use the following
equalities which are obtained by contracting Eqs. (4.11) and (4.12) with $%
k_1^\mu $ 
\begin{equation}
k_1^\mu k_2^\nu [T_{\mu \nu }^{(1)ab}+T_{\mu \nu }^{(2)ab}]=-g^2f^{abc}%
\overline{v}(p_2)\frac{\lambda ^c}2{\bf k}_1u(p_1)  \eqnum{4.18}
\end{equation}
and 
\begin{equation}
k_1^\mu k_2^\nu T_{\mu \nu }^{(3)ab}=g^2f^{abc}\overline{v}(p_2)\frac{%
\lambda ^c}2{\bf k}_1u(p_1)+m^2\widetilde{S}^{ab}.  \eqnum{4.19}
\end{equation}
These equalities and Eq. (4.15) lead to 
\begin{equation}
k_1^\mu k_2^\nu T_{\mu \nu }^{ab}=m^2S^{ab}.  \eqnum{4.20}
\end{equation}
This result allows us to write the last term in Eq. (4.1) in the form 
\begin{equation}
T_{\mu \nu }^{ab}T_{\mu ^{\prime }\nu ^{\prime }}^{ab*}Q^{\mu \mu ^{\prime
}}(k_1)Q^{\nu \nu ^{\prime }}(k_2)=S^{ab}S^{ab*}.  \eqnum{4.21}
\end{equation}

Substituting Eqs. (4.10), (4.17) and (4.21) in Eq. (4.1), we arrive at 
\begin{equation}
2ImT_1=\frac 12\int d\tau T_{\mu \nu }^{ab}T_{\mu ^{\prime }\nu ^{\prime
}}^{ab*}P^{\mu \mu ^{\prime }}(k_1)P^{\nu \nu ^{\prime }}(k_2)+\frac 12\int
d\tau S^{ab}S^{ab*}.  \eqnum{4.22}
\end{equation}

The ghost diagram in Fig. (8f) can be given by folding the tree diagram in
Fig. (9d) with its conjugate. Therefore, the imaginary part of Fig. (8f), by
the Feynman rules, can be written as 
\begin{equation}
2ImT_2=-\int d\tau S^{ab}S^{ab*}  \eqnum{4.23}
\end{equation}
In complete analogy with the two-boson scattering discussed in the preceding
section, the second term in Eq. (4.22) can only cancel half of the above
amplitude. The reason for this still is due to the difference between the
symmetry factors of Figs. (8e) and (8f). To achieve a complete cancellation,
it is necessary to consider the contribution of the diagram in Fig. (8g).
This diagram can also be treated as a limit of the diagram in Fig. (8e) when
the momentum of one internal line in the loop tends to infinity, 
\begin{equation}
\begin{array}{c}
2ImT_3=\frac 12\int d\tau {\lim_{\left| k_{2\mu }\right| \rightarrow \infty }%
}T_{\mu \nu }^{(3)ab}T_{\mu ^{\prime }\nu ^{\prime }}^{(3)ab*}[P^{\mu \mu
^{\prime }}(k_1)P^{\nu \nu ^{\prime }}(k_2) \\ 
+Q^{\mu \mu ^{\prime }}(k_1)g^{\nu \nu ^{\prime }}+g^{\mu \mu ^{\prime
}}Q^{\nu \nu ^{\prime }}(k_2)-Q^{\mu \mu ^{\prime }}(k_1)Q^{\nu \nu ^{\prime
}}(k_2)].
\end{array}
\eqnum{4.24}
\end{equation}
In the limit:$\left| k_{2\mu }\right| \rightarrow \infty $, comparing to the
second terms in Eqs. (4.7), (4.12) and (4.19), the first terms in these
equations can be ignored. Thus, Eqs. (4.7), (4.12) and (4.19) respectively
reduce to 
\begin{equation}
k_1^\mu T_{\mu \nu }^{(3)ab}\approx k_{2\nu }S^{ab}.  \eqnum{4.25}
\end{equation}
\begin{equation}
k_2^\mu T_{\mu \nu }^{(3)ab}\approx k_{1\mu }\widetilde{S}^{ab}  \eqnum{4.26}
\end{equation}
and 
\begin{equation}
k_1^\mu k_2^\nu T_{\mu \nu }^{(3)ab}\approx M^2\widetilde{S}^{ab} 
\eqnum{4.27}
\end{equation}
On inserting these expressions into Eq. (4.24), we have 
\begin{equation}
\begin{array}{c}
2ImT_3=\frac 12\int d\tau {\lim_{\left| k_{2\mu }\right| \rightarrow \infty }%
}T_{\mu \nu }^{(3)ab}T_{\mu ^{\prime }\nu ^{\prime }}^{(3)ab*}P^{\mu \mu
^{\prime }}(k_1)P^{\nu \nu ^{\prime }}(k_2) \\ 
+\frac 12\int d\tau S^{ab}S^{ab*}.
\end{array}
\eqnum{4.28}
\end{equation}
Thus, as shown before, it is indeed possible to find a way which allows us
to isolate from Fig. (8g) the unphysical part of the amplitude like the
second term in Eq. (4.28).

Summing the results denoted in Eqs .(4.22), (4.23) and (4.28), we finaly
obtain the imaginary part of the total amplitude such that 
\begin{equation}
\begin{array}{c}
2ImT=\frac 12\int d\tau T_{\mu \nu }^{ab}T_{\mu ^{\prime }\nu ^{\prime
}}^{ab*}P^{\mu \mu ^{\prime }}(k_1)P^{\nu \nu ^{\prime }}(k_2) \\ 
+\frac 12\int d\tau {\lim_{\left| k_{2\mu }\right| \rightarrow \infty }}%
T_{\mu \nu }^{\left( 3\right) ab}T_{\mu ^{\prime }\nu ^{\prime }}^{\left(
3\right) ab*}P^{\mu \mu ^{\prime }}(k_1)P^{\nu \nu ^{\prime }}(k_2)
\end{array}
\eqnum{4.29}
\end{equation}
in which the unphysical contributions are all cancelled. Thus, the unitarity
is ensured.

\section{Comments and discussions}

In the previous sections, the unitarity of our theory has been illustrated
by evaluating the imaginary parts of two-gauge boson and fermion-antifermion
scattering amplitudes up to the fourth order perturbation. The imaginary
parts of the amplitudes were calculated by means of the L-C rule. In this
kind of calculation, we have to first work in the Feynman gauge because the
formula used requires the intermediate states to be complete. The gauge
boson propagator given in the Feynman gauge contains unphysical longitudinal
intermediate excitations though, it has been proved that these unphysical
intermediate states are eventually cancelled in the S-matrix elements,
leaving only physical transverse intermediate states in the S-matrix
elements as given by the gauge boson propagator written in the unitary gauge 
\begin{equation}
D_{\mu \nu }(k)=\frac{g_{\mu \nu }-k_\mu k_\nu /m^2}{k^2-m^2+i\varepsilon }.
\eqnum{5.1}
\end{equation}
Although the unitarity of the S-matrix elements is proved in the Feynman
gauge, it would be true for other gauges since it has been exactly proved
that the S-matrix is gauge-independent.

As mentioned in Introduction, In the previous works of examining the
unitarity of some kinds of massive non-Abelian gauge field theories$%
^{[14][15]}$, the gauge boson propagator in the Landau gauge 
\begin{equation}
D_{\mu \nu }(k)=\frac{g_{\mu \nu }-k_\mu k_\nu /k^2}{k^2-m^2+i\varepsilon } 
\eqnum{5.2}
\end{equation}
was chosen to calculate the imaginary part of scattering amplitudes by the
L-C rule. For such a calculation, as mentioned in Introduction, the Landau
gauge actually is not suitable since the intermediate states characterized
by the transverse projector appearing in the numerator of the above
propagator does not form a complete set. The unsuitability of the procedure
may be seen from the massless gauge theory. The unitarity of the theory was
exampled by computing the imaginary part of the fermion-antifermion
scattering amplitude of order $g^4$ in the Feynman gauge$^{[19]}$. However,
if one tries to perform the proof in the Landau gauge, he could not get a
reasonable result. Particularly, the longitudinal projector $k_\mu k_\nu
/k^2 $ in Eq. (5.2) where $m=0$ could not be given an unambiguous definition
on the mass shell $k^2=0$ since the momentum $k$ on the mass shell becomes
an isotropic vector.

Another point we would like to emphasize is that for examining the unitarity
of a massive gauge field theory, it is only necessary to evaluate the
S-matrix element between the physical transversely polarized states of gauge
bosons. In this way, it was shown in sections (2) and (3) that the unitarity
is well satisfied. Particularly, the calculation in section 3 indicates that
in all the diagrams depicted in Figs.(3) and (4) except for the loop
diagrams involving Fig. (3k) and a part of Fig. (4e), there is a natural
cancellation among the contributions coming from the unphysical gauge boson
and ghost particle intermediate states, without the help of any scalar
particle. This result and the theoretical logic strongly suggest that the
same cancellation between the unphysical contributions arising from Fig.
(3k) and a part of Fig. (4e) is definite to happen. To achieve this
cancellation, the loop diagram formed by the gauge boson four-line vertex is
necessary to be recast in the form as if it is given by the two-particle
intermediate states so as to be able to compare its contribution with that
given by the other loop diagrams. For this purpose, we proposed in section 3
a reasonable limiting procedure which allows us to reach the cancellation
mentioned above. The results we obtained are undoubtedly correct. It should
be noted that in all the previous investigations$^{[12][14][15]}$ of the
unitarity problem, the loop diagrams given by the gauge boson four-line
vertex such as Figs. (3k) and (8g) were never taken into account in the
cancellation of the unphysical amplitudes. From the calculations described
in sections (3) and (4), it is clearly seen that these loop diagrams play an
essential role to guarantee the cancellation of the unphysical part of the
amplitudes and hence the unitarity of the S-matrix elements. It is mentioned
that the theories presented in Refs. (2) and (6) which were pointed out to
be non-unitary by Mohapatra et. al. and some others$^{[4,12.15]}$ are not
correct because the Feynman rule concerning the closed ghost loop has an
extra factor $\frac 12$ other than $1$ as given in our theory. In this
paper, the unitarity has been proved in the perturbation approximation up to
the order of $g^4$. For higher order approximations, we believe that for a
given process, if all the diagrams are taken into account and treated
appropriately, the unitarity would be proved to be no problem.

It should be noted that in this paper and the former papers, we only limit
ourself to discuss the theory ( for example, the QCD with massive gluons) in
which all the gauge bosons are required to have the same masses. This
requirement is necessary to make the theory to be gauge-invariant and
unitary just as the case we met in nuclear physics where the nucleon-pion
interacting system is of SU(2)-symmetry provided that the masses of all the
pions are considered to be the same and the mass difference between proton
and neutron is ignored..Similarly, for the weak interaction theory in which
apart from the vector currents of fermions, the axial vector currents of
fermions are included as well, certainly, we may build up a SU(2)- symmetric
theory without introducing the Higgs mechanism. The action of this theory is
constrained by the Lorentz condition and may contain the gauge boson mass
term in it. But, the gauge-invariance of the theory requires that all the
gauge bosons must be of the same mass and the fermions must be massless. In
this case, the unitarity of the theory, as easily checked, is no problem.
Nevertheless, if the charged and neutral gauge bosons are required to have
different masses and charged fermions are massive, it is inevitable to work
with the theory constructed by means of the Higgs mechanism.

\section{Acknowledgment}

This project was supported in part by National Natural Science Foundation of
China.

\section{Appendix: Examination of signs of the imaginary parts of the loop
diagrams{\bf \ }}

It was mentioned in section (3) that the imaginary part of the matrix
element of the loop in Fig. (7a) has the same sign as that given by the loop
in Fig. (7b) at least for the large momentum $k_2$. To convince ourself of
this point, we investigate the imaginary part of the loops in a parametric
representation. In this representation, the propagator will be expressed in
the form $^{[17]}$%
\begin{equation}
\frac 1{k_i^2-m^2+i\varepsilon }=-i{\int_0^\infty }d\alpha _ie^{i\alpha
_i(k_i^2-m^2+i\varepsilon )}.  \eqnum{A1}
\end{equation}
With this representation, the matrix element shown in Eq. (3.81) for Fig.
(7a) may be rewritten as 
\begin{equation}
\Pi _{\lambda \lambda ^{\prime }}^{(1)ab}(q)=-3g^2f^{abc}f^{bcd}g_{\lambda
\lambda ^{\prime }}J^{(1)}  \eqnum{A2}
\end{equation}
where 
\begin{equation}
\begin{array}{c}
J^{(1)}=\int \frac{d^4k}{(2\pi )^4}\frac 1{k^2-m^2+i\varepsilon } \\ 
=-i{\int_0^\infty }d\alpha \int \frac{d^4k}{(2\pi )^4}e^{i\alpha
(k^2-m^2+i\varepsilon )} \\ 
=-\frac 1{(4\pi )^2}{\int_0^\infty }\frac{d\alpha }{\alpha ^2}e^{-i\alpha
(m^2-i\varepsilon )}.
\end{array}
\eqnum{A3}
\end{equation}
In the above, we have used the representation in Eq. (A1) and the formula of
Fresnel integral$^{[17]}$%
\begin{equation}
\int \frac{d^4k}{(2\pi )^4}e^{i\alpha k^2}=\frac{-i}{(4\pi \alpha )^2} 
\eqnum{A4}
\end{equation}

For the loop in Fig. (7b), we confine ourselves to investigate its
expression in the limit of large momentum $k_2$ for the purpose of comparing
to the one given in Eqs. (A2) and (A3). As stated in Eqs. (3.85)-(3.88), in
order to convert the matrix element of Fig.(7b) to the one for Fig.(7a), it
is necessary to take the approximate expression denoted in Eq. (3.85) and
the limit assumed in Eq. (3.86) which is applied to Eq. (3.85). To achieve
such a conversion, as easily verified, instead of Eq. (3.85), we may simply
take an equivalent approximate expression such that 
\begin{equation}
g^{\mu \mu ^{\prime }}g^{\nu \nu ^{\prime }}\Gamma _{\mu \nu \lambda
}(k_1,k_2,q)\Gamma _{\mu ^{\prime }\nu ^{\prime }\lambda ^{\prime
}}(k_1,k_2,q)\rightarrow 6k_{2\lambda }k_{2\lambda ^{\prime }}.  \eqnum{A5}
\end{equation}
With this expression, Eqs. (3.83) and (3.84) may be rewritten as 
\begin{equation}
\Pi _{\lambda \lambda ^{\prime }}^{(2)ab}(q)\approx
3g^2f^{acd}f^{bcd}J_{\lambda \lambda ^{\prime }}^{(2)}(q)  \eqnum{A6}
\end{equation}
where 
\begin{equation}
\begin{array}{c}
J_{\lambda \lambda ^{\prime }}^{(2)}(q)=\int \frac{d^4k_1}{(2\pi )^4}\frac{%
d^4k_2}{(2\pi )^4}(2\pi )^4\delta ^4(q-k_1-k_2) \\ 
\times \frac{k_{2\lambda }k_{2\lambda ^{\prime }}}{(k_1^2-m^2+i\varepsilon
)(k_2^2-m^2+i\varepsilon )}.
\end{array}
\eqnum{A7}
\end{equation}
Employing the parametrization given in Eq. (A1) and the Fourier
representation of the $\delta $-function, Eq. (A7) reads 
\begin{equation}
\begin{array}{c}
J_{\lambda \lambda ^{\prime }}^{(2)}(q)={\int_0^\infty }d\alpha _1d\alpha
_2\int d^4z\int \frac{d^4k_1}{(2\pi )^4}e^{i\alpha _1k_1^2-\frac i4\frac{z^2%
}{\alpha _1}-iqz} \\ 
\times \partial _\lambda ^z\partial _{\lambda ^{\prime }}^z\int \frac{d^4k_2%
}{(2\pi )^4}e^{i\alpha _2k_2^2-\frac i4\frac{z^2}{\alpha _2}-i(\alpha
_1+\alpha _2)(m^2-i\varepsilon )}.
\end{array}
\eqnum{A8}
\end{equation}
Upon completing the integrations over $k_1,k_2$ and $z$ by using the formula
of Fresnel integral, we have 
\begin{equation}
\begin{array}{c}
J_{\lambda \lambda ^{\prime }}^{(2)}(q)=\frac i{(4\pi )^2}{\int_0^\infty }%
\frac{d\alpha _1d\alpha _2}{(\alpha _1+\alpha _2)^2}[g_{\lambda \lambda
^{\prime }}\frac i{2\alpha _2}-\frac 1{4\alpha _2^2}\partial _\lambda
^q\partial _{\lambda ^{\prime }}^q] \\ 
\times e^{iQ(q,\alpha )}
\end{array}
\eqnum{A9}
\end{equation}
where 
\begin{equation}
Q(q,\alpha )=\frac{\alpha _1\alpha _2}{\alpha _1+\alpha _2}q^2-(\alpha
_1+\alpha _2)(m^2-i\varepsilon ).  \eqnum{A10}
\end{equation}
Inserting the identity 
\begin{equation}
{\int_0^\infty }dx\delta (x-\alpha _1-\alpha _2)=1  \eqnum{A11}
\end{equation}
into Eq. (A9) and then making the transformation $\alpha _i\rightarrow
x\alpha _i$, one can write 
\begin{equation}
\begin{array}{c}
J_{\lambda \lambda ^{\prime }}^{(2)}(q)=\frac i{(4\pi )^2}{\int_0^1}d\alpha
_1d\alpha _2\delta (1-\alpha _1-\alpha _2) \\ 
\times {\int_0^\infty }\frac{dx}x(\frac i{2x}g_{\lambda \lambda ^{\prime
}}+\alpha _1^2q_\lambda q_{\lambda ^{\prime }})e^{ixQ(q,\alpha )}.
\end{array}
\eqnum{A12}
\end{equation}
Noticing the equalities shown in Eqs. (3.29) and (4.14), the second term in
the parenthesis, actually, can be ignored in the scattering amplitude. Thus,
the function $J_{\lambda \lambda ^{\prime }}^{(2)}(q)$ is only proportional
to the unit tensor $g_{\lambda \lambda ^{\prime }}$. This result precisely
justifies the limit taken in Eq. (3.86). On substituting Eq. (A12) into Eq.
(A6), and performing the integration over $\alpha _2$, one gets 
\begin{equation}
\Pi _{\lambda \lambda ^{\prime }}^{(2)ab}(q)=-3g^2f^{acd}f^{bcd}g_{\lambda
\lambda ^{\prime }}J^{(2)}(q)  \eqnum{A13}
\end{equation}
where 
\begin{equation}
J^{(2)}(q)=\frac 1{(4\pi )^2}{\int_0^1}d\alpha {\int_0^\infty }\frac{dx}{2x^2%
}e^{ixQ(q,\alpha )}  \eqnum{A14}
\end{equation}
in which 
\begin{equation}
Q(q,\alpha )=\alpha (1-\alpha )q^2-m^2.  \eqnum{A15}
\end{equation}

Now, we are in position to examine the imaginary parts of the functions $\Pi
_{\lambda \lambda ^{\prime }}^{(1)ab}$ and $\Pi _{\lambda \lambda ^{\prime
}}^{(2)ab}$. We First write down the imaginary parts of the functions $%
J^{(1)}$ and $J^{(2)}$, 
\begin{equation}
ImJ^{(1)}=\frac 1{(4\pi )^2}{\int_0^\infty }\frac{dx}{x^2}\sin (xm^2) 
\eqnum{A16}
\end{equation}
and 
\begin{equation}
ImJ^{(2)}=\frac 1{(4\pi )^2}{\int_0^1}d\alpha {\int_0^\infty }\frac{dx}{2x^2}%
\sin \{x[\alpha (1-\alpha )q^2-m^2]\}.  \eqnum{A17}
\end{equation}
For the integral over $x$, obviously, the major contribution arises from the
integrand at the neighborhood of the origin. Therefore 
\begin{equation}
ImJ^{(1)}\geq 0.  \eqnum{A.18}
\end{equation}
As for the $ImJ^{(2)}$, the integral over $\alpha $ may be estimated by
taking the mean value $\frac 12$ of the variable $\alpha $ in the integrand.
Thus, 
\begin{equation}
ImJ^{(2)}\approx \frac 1{(4\pi )^2}{\int_0^\infty }\frac{dx}{2x^2}\sin [x(%
\frac 14q^2-m^2)].  \eqnum{A19}
\end{equation}
It is well known that in the physical region, 
\begin{equation}
q^2\geq 4m^2  \eqnum{A20}
\end{equation}
where $q^2=4m^2$ is the starting point of a cut which is the solution of the
following Landau equations$^{[17]}$%
\begin{equation}
\begin{array}{c}
\lambda _1(k^2-m^2)=0, \\ 
\lambda _2[(q-k)^2-m^2]=0, \\ 
\lambda _1k_\mu -\lambda _2(q-k)_\mu =0.
\end{array}
\eqnum{A21}
\end{equation}
In view of Eq. (A20), we may conclude 
\begin{equation}
ImJ^{(2)}\geq 0.  \eqnum{A22}
\end{equation}
The results in Eqs. (A18) and (A22) straightforwardly lead to that the
imaginary parts of the $\Pi _{\lambda \lambda ^{\prime }}^{(1)ab}(q)$ and $%
\Pi _{\lambda \lambda ^{\prime }}^{(2)ab}(q)$ have the same sign as we see
from Eq. (A2) and (A13).

At last , we would like to mention the imaginary part of the loop in Fig.
(7c). The expression of the loop is 
\begin{equation}
\Pi _{\lambda \lambda ^{\prime }}^{(3)ab}(q)=-g^2f^{acd}f^{bcd}J_{\lambda
\lambda ^{\prime }}^{(3)}(q)  \eqnum{A23}
\end{equation}
where 
\begin{equation}
J_{\lambda \lambda ^{\prime }}^{(3)}(q)=-\int \frac{d^4k_1}{(2\pi )^4}\frac{%
d^4k_2}{(2\pi )^4}(2\pi )^4\delta ^4(q-k_1-k_2)\frac{k_{1\lambda
}k_{2\lambda ^{\prime }}}{(k_1^2-m^2+i\varepsilon )(k_2^2-m^2+i\varepsilon )}%
.  \eqnum{A24}
\end{equation}
Completely following the procedure formulated in Eqs. (A8)-(A12), one may
derive 
\begin{equation}
J_{\lambda \lambda ^{\prime }}^{(3)}(q)=\frac i{(4\pi )^2}{\int_0^1}d\alpha {%
\int_0^\infty }\frac{dx}x[\frac i{2x}g_{\lambda \lambda ^{\prime }}-\alpha
(1-\alpha )q_\lambda q_{\lambda ^{\prime }}]e^{ixQ(q,\alpha )}.  \eqnum{A25}
\end{equation}
Neglecting the second term containing $q_\lambda q_{\lambda ^{\prime }}$ and
then substituting the above expression into Eq. (A23), we can write 
\begin{equation}
\Pi _{\lambda \lambda ^{\prime }}^{(3)ab}(q)=-g^2f^{acd}f^{bcd}g_{\lambda
\lambda ^{\prime }}J^{(3)}(q)  \eqnum{A26}
\end{equation}
where 
\begin{equation}
J^{(3)}(q)=-\frac 1{(4\pi )^2}{\int_0^1}d\alpha {\int_0^\infty }\frac{dx}{%
2x^2}e^{ix[\alpha (1-\alpha )q^2-m^2]}.  \eqnum{A27}
\end{equation}
Clearly, the sign of the imaginary part 
\begin{equation}
ImJ^{(3)}(q)=-\frac 1{(4\pi )^2}{\int_0^1}d\alpha {\int_0^\infty }\frac{dx}{%
2x^2}\sin \{x[\alpha (1-\alpha )q^2-m^2]\}  \eqnum{A28}
\end{equation}
is opposite to the $ImJ^{(2)}(q)$ shown in Eq. (A17). Therefore, the
imaginary parts of the $\Pi _{\lambda \lambda ^{\prime }}^{(3)ab}(q)$ and
the $\Pi _{\lambda \lambda ^{\prime }}^{(2)ab}(q)$ have opposite signs.

\section{References}

\begin{itemize}
\item[1]  J. C. Su, IL Nuovo Cimento, {\bf 117 B} 203 (2002).
\end{itemize}

\begin{enumerate}
\item[2]  T. Kunimasa and T. Goto, Prog. Theor. Phys. {\bf 37}, 452 (1967).

\item[3]  M. Veltman, Nucl. Phys. B7, 637 (1968).

\item[4]  J. Reiff and M.Veltman, Nucl. Phys. {\bf B 13}, 545 (1969).

\item[5]  E. Fradkin and I. Tyutin, Phys. Lett.{\bf \ B 30}, 562 (1969).

\item[6]  G. Curci and R. Ferrari, Nuovo Cimento {\bf 32 A}, 111 (1971).

\item[7]  J. P. Hsu and E. C. G. Sudarshan, Phys. Rev.{\bf \ D 9}, 1678
(1974).

\item[8]  R. Delbourgo and G. Thompson, Phys. Rev. Lett. {\bf 57}, 2610
(1986).

\item[9]  G. t 'Hooft, Nucl.Phys. {\bf B 35}, 167 (1971).

\item[10]  C.Llewellyn-Smith, Phys. Lett. {\bf 46 B}, 233 (1973).
\end{enumerate}

\begin{enumerate}
\item[11]  J. Cornwall, D. Levin and G. Tikttopoulos, Phys. Rev. Lett. {\bf %
30}, 1268 (1973); Phys. Rev. {\bf D 10}, 1145 (1974).

\item[12]  R. N. Mohapatra, S . Sakakibara and J. Sucher, Phys. Rev. {\bf D
10}, 1844 (1974).

\item[13]  D. Kosinski and L. Szymanowski, Phys. Rev. Lett. {\bf 58}, 2001
(1987).

\item[14]  J. Kubo, Phys. Rev. Lett. {\bf 58}, 2000 (1987).

\item[15]  R. Delbourgo, S. Twisk and G. Thompson, Int. J. Mod. Phys. {\bf A
3}, 435 (1988).

\item[16]  A. Burnel, Phys. Rev. D33, 2981 (1986); ibid. {\bf D 33}, 2985
(1986).

\item[17]  C. Itzykson and F-B. Zuber, Quantum Field Theory, McGraw-Hill,
New York (1980).

\item[18]  R. E.Cutkosky, J. Math. Phys. {\bf 1}, 429 (1960).

\item[19]  Ta-Pei Cheng and Ling-Fong Li, Gauge Theory of Elementary
Particle Physis, Clarendon Press, Oxford (1984).
\end{enumerate}

\begin{description}
\item[20]  M. E. Peskin and D. V. Schoeder, An Introduction to Quantum Field
Theory, Addison-Wesley Publishing Company, New York (1995).
\end{description}


\begin{references}
\bibitem{}  FIGURE CAPTION

Fig. (1): The tree diagram for fermion and antifermion scattering.

Fig. (2): The tree diagram for two-gauge boson scattering.

Fig. (3): The fourth-order diagrams for two-gauge boson scattering with only
the gauge boson intermediate states.

Fig. (4): The fourth-order diagrams for two-gauge boson scattering with only
the ghost particle intermediate states.

Fig. (5): The tree diagrams used to give all the diagrams in Fig. 3 through
folding them with their conjugates

Fig. (6): The tree diagrams used to give all the diagrams in Fig. 4 through
folding them with their conjugates.

Fig. (7a): The one-loop diagram formed by four-line gauge boson vertex.

Fig. (7b): The one-loop diagram formed by three-line gauge boson vertex.

Fig. (7c): The one-loop diagram formed by ghost intermediate states.

Fig. (8): The fourth-order diagrams for fermion-antifermion scattering.

\begin{center}
Fig. (9): The tree diagrams used to give all the diagrams in Fig. 8 through
folding them with their conjugates.
\end{center}
\end{references}
\end{document}